\documentclass[%
 reprint,
superscriptaddress,
 amsmath,amssymb,amsfonts,
 aps,
pra,
 floatfix,
 longbibliography,nobibnotes
]{revtex4-2}

\usepackage[utf8]{inputenc}

\usepackage[pdftex]{graphicx}
\graphicspath{{main}}
\usepackage{float} 
\usepackage{xcolor}
\usepackage{verbatim}
\usepackage{soul}
\usepackage[normalem]{ulem}
\usepackage[caption=false]{subfig}

\usepackage[pdftex,colorlinks=true]{hyperref}
\hypersetup{
  colorlinks=true,
  linkcolor=blue,
  urlcolor=cyan,
}

\usepackage{bm}

\usepackage{mathtools}

\DeclarePairedDelimiterX\braket[2]{\langle}{\rangle}{#1 \delimsize\vert #2}
\DeclarePairedDelimiterX\expval[3]{\langle}{\rangle}{#1 \delimsize\vert #2  \delimsize\vert #3}
\DeclarePairedDelimiterX\proj[2]{\delimsize\vert#1\rangle}{\langle#2\delimsize\vert}{ }

\newcommand{\vect}[1]{\boldsymbol{\mathbf{#1}}}

\usepackage{microtype}

\begin{document}

\title{Universal Spin Squeezing Dynamical Phase Transitions across Lattice Geometries, Dimensions, and Microscopic Couplings}

\author{Arman Duha}
\affiliation{Department of Physics, Oklahoma State University, Stillwater, Oklahoma 74078, USA}

\author{Thomas Bilitewski}
\email{thomas.bilitewski@okstate.edu}
\affiliation{Department of Physics, Oklahoma State University, Stillwater, Oklahoma 74078, USA}

\date{\today}

\begin{abstract}
Recent work has identified a dynamical squeezing phase transition in power-law interacting bilayer XXZ spin models, separating a fully collective phase with Heisenberg-limited squeezing from a partially collective phase with universal critical scaling. %
Here we test and establish the universality of this transition along two qualitatively different microscopic axes: lattice geometry, by studying square, triangular, and honeycomb $2\mathrm{D}$ bilayers as well as $1\mathrm{D}$ ladders, and a symmetry-preserving rescaling $\lambda$ of the interlayer couplings relative to the intralayer ones. %
Combining a Bogoliubov instability analysis with discrete truncated Wigner simulations, we find that the transition persists across all four lattice geometries and over a wide range of $\lambda$ with critical exponents consistent within error, providing strong evidence for a genuine non-equilibrium universality class. 
The Bogoliubov theory recovers the previously identified scaling $a_Z^* \propto L$ in the long-range interacting regime $\alpha < d+2$, and yields an analytical scaling $a_Z^* \propto L^{2/(\alpha-d)}$ for the critical aspect ratio with system size for $\alpha>d+2$, with $\alpha$ the power-law exponent in dimension $d$. This uncovers a previously unrecognized sub-linear regime for short-range interactions. %
 By tuning $\lambda$ we vary the interlayer coupling strength at fixed layer spacing, demonstrating that the dynamical transition can be driven purely through interaction engineering without modifying the underlying geometry. %
These findings provide a versatile route toward controlling entanglement generation in Rydberg-array, polar molecule, and trapped-ion platforms with applications in quantum sensing and simulation.
\end{abstract}
\maketitle

\section{Introduction}%
Long-range interacting quantum spin systems have emerged as versatile platforms for the controlled study of many-body nonequilibrium dynamics \cite{RevModPhys.95.035002}. Realized in Rydberg atom arrays \cite{Browaeys2020,Saffman2010,Morgado2021}, polar molecules \cite{Baranov2012,Bohn2017,Moses2017}, magnetic atoms \cite{Chomaz2023}, and trapped ions \cite{Blatt2012,Monroe2021} and cavities \cite{Norcia2018,Davis2019,Mivehvar2021}, these systems realize power-law interactions, with exponents ranging from infinite-range cavity-mediated couplings to short-range van der Waals interactions, with the ability to control individual constituents at the single-particle level. The lattice geometry of these platforms is itself a tunable degree of freedom; optical lattices and reconfigurable tweezer arrays enable the realization of bilayer geometries built from one-dimensional ladders or two-dimensional layers \cite{Du2024,Hawaldar2024}, or three-dimensional structures \cite{Barredo2018,PhysRevLett.130.180601}. The interactions  can be reshaped through Floquet engineering, where periodic pulse sequences transform the native couplings of a given platform into target Hamiltonians with engineered spin structure \cite{Lukin_2020_Robust,Geier2021,Scholl2022,Christakis2023,Miller2024,PRXQuantum.4.010334}. Together, geometric and interaction engineering provide a rich and growing toolbox for designing nonequilibrium dynamics in long-range quantum spin systems.

A particularly important application of this toolbox is the dynamical generation of metrologically useful entanglement \cite{RevModPhys.89.035002,RevMod_Metrology_2018,montenegro2024reviewquantummetrologysensing} via time evolution under an interacting Hamiltonian from initial unentangled product states. Spin-squeezed states \cite{Kitagawa1993,Wineland1992,Wineland1994}, with quantum noise reduced below the standard quantum limit, are one such example. Spin squeezing can provide direct sensitivity gains in atomic clocks \cite{pedrozo2020entanglement,Eckner2023} and magnetometers \cite{PhysRevLett.104.013601,PhysRevLett.109.253605,PhysRevLett.113.103004} and has long been a central target for entanglement generation in atomic platforms. One-axis twisting \cite{Kitagawa1993}, one paradigmatic mechanism to generate spin squeezing, relies on infinite-range interactions to coherently squeeze a collective spin and achieves a sensitivity improvement over using unentangled states scaling as $1/N^{2/3}$ in the number of spins $N$, while two-axis twisting \cite{Kitagawa1993} reaches the ultimate Heisenberg limit $1/N$. %

Finite-range systems with power-law interactions have also been predicted to generate scalable spin squeezing \cite{begg2026scalable,fossfeig2016entanglementspinsqueezinginfiniterangeinteractions,PhysRevLett.125.223401,PhysRevLett.126.113401,Block_2024,Roscilde2022,Roscilde2023,Roscilde2024,PhysRevA.109.L061304,koyluoglu2025squeezingheisenberglimitlocally}, which was recently demonstrated in a number of experiments \cite{Eckner2023,Franke2023a,Bornet2023a,Hines2023}. One particular insight explored in these works is the connection between scalable spin squeezing and equilibrium phases of matter \cite{begg2026scalable,PhysRevLett.125.223401,Block_2024,Roscilde2024}. Specifically, scalable squeezing has been linked with dynamics thermalizing to the easy-plane ferromagnetic phase \cite{Block_2024} and this picture was then extended to the case of quasi-long range ordered phases \cite{Roscilde2024}, and to disordered systems \cite{kaplanlipkin2025theoryscalablespinsqueezing}, and most recently to magnetic ordering and percolation on generic graph geometries \cite{solfanelli2026robustspinsqueezingquantumnetworks}.

 Scalable squeezing was further connected to non-equilibrium universality \cite{7g55-lpff}, where a nonequilibrium squeezing phase transition was identified between two distinct dynamical regimes: a fully collective phase with Heisenberg-limited squeezing, and a partially collective phase with scalable, but sub-Heisenberg, squeezing. %
 However, several questions remain open. 
 Whether the universality persists across lattice geometries, across general values of $\alpha$, and across other symmetry-preserving microscopic variations of the Hamiltonian, is an open question. The role of aspect ratio $a_Z/L$ as the natural control parameter of the transition was established empirically without an analytical understanding that generalizes beyond the specific model considered. Finally, the fate of the transition and the validity of this control parameter for short-range-dominated interactions ($\alpha>d+2$) remained unexplored.

In this paper we address these unresolved questions. We establish the universality of the transition across lattice geometries for a range of $\alpha$, including the experimentally relevant case of $\alpha = 3$. We further introduce an independent microscopic control parameter, $\lambda$, that rescales the strength of all interlayer couplings relative to the intralayer ones. In both cases, the same critical exponents are found, strongly suggesting a genuine nonequilibrium universality class. %
Additionally, we establish that tuning $\lambda$ drives the system across the dynamical transition without any change in the underlying geometry, a practically significant advantage given that the layer spacing is typically fixed. %
The qualitative agreement between the analytical Bogoliubov theory and direct discrete truncated Wigner (dTWA) simulations across all lattice geometries, dimensions, and values of $\lambda$ considered here establishes the Bogoliubov analysis as a reliable and computationally inexpensive method to predict the phase boundary. Building on this, we obtain an analytical understanding of when the aspect ratio $a_Z/L$ is the correct control parameter by analyzing the Bogoliubov dispersion relations. This recovers the scaling $a_Z^* \propto L$ for the long-range $\alpha < d+2$ cases discussed in Ref.~\cite{7g55-lpff} and predicts a previously unrecognized change in scaling $a_Z^* \propto L^{2/(\alpha - d)}$ for short-range $\alpha > d+2$, which we confirm within the numerical dTWA results.

The remainder of the paper is organized as follows. In Sec.~\ref{sec:model} we introduce the model, the Holstein-Primakoff transformation underlying the Bogoliubov analysis, the dTWA simulations, and the variance-scaling criterion that connects the dynamical phases to their metrological utility. In Sec.~\ref{sec:phase} we map out the phase boundary as a function of lattice geometry, dimensionality, and $\lambda$. In Sec.~\ref{sec:scaling_aZL} we derive and verify the analytical scaling of $a_Z^*$ with system size. In Sec.~\ref{sec:universality} we test the universality of the partially collective phase across lattice geometries and values of $\lambda$. We close with an outlook in Sec.~\ref{sec:outlook}.

\section{Model}\label{sec:model}
We consider power-law interacting spin-$1/2$ systems in bilayer geometries, described by the Hamiltonian
\begin{equation}
\hat{H} = \frac{1}{2} \sum_{\eta} \sum_{\boldsymbol{i},\boldsymbol{j} \in \eta} V_{\boldsymbol{i}\boldsymbol{j}} \, \vec{s}_{\boldsymbol{i}} \cdot \vec{s}_{\boldsymbol{j}} +  \lambda \sum_{\mathclap{\boldsymbol{i} \in A, \boldsymbol{j} \in B}} \, V_{\boldsymbol{i}\boldsymbol{j}} \, (\hat{s}_{\boldsymbol{i}}^x \hat{s}_{\boldsymbol{j}}^x +\hat{s}_{\boldsymbol{i}}^y \hat{s}_{\boldsymbol{j}}^y ) \label{eq:model}
\end{equation}
where $\hat{s}^\mu_i = \hat{\sigma}^\mu_i/2$ are spin-$1/2$ operators acting on site $i$, and $\eta = A,\,B$ labels the two layers. The first term describes Heisenberg interactions within each layer, while the second describes XX exchange between layers, scaled by a dimensionless ratio $\lambda$ that controls the relative strength of interlayer to intralayer coupling. The two layers are separated by a distance $a_Z$, measured in units of the in-layer spacing $a_{\mathrm{lat}} = 1$, and have linear extent $L$, with $N = L^d$ spins per layer, with $d=1$ (ladder) or $d=2$ (bilayer). 

The interactions depend only on the distance between sites and decay as a power-law, $V_{ij} = |\vect{r}_i - \vect{r}_j|^{-\alpha}$. The exponent $\alpha$ controls the range of the interactions and is set by the underlying physical system. This covers a variety of experimental platforms \cite{RevModPhys.95.035002} such as Rydberg atoms ($\alpha=3,6$) \cite{Browaeys2020,Saffman2010,Morgado2021}, polar molecules ($\alpha=3$)\cite{Baranov2012,Bohn2017,Moses2017}, magnetic atoms ($\alpha=3$)\cite{Chomaz2023}, trapped ions ($0<\alpha<3$) \cite{Blatt2012, Monroe2021}, or cavity systems ($\alpha=0$) \cite{Norcia2018,Davis2019,Mivehvar2021}.

The spin structure of interactions in Eq.~\eqref{eq:model}, with intralayer Heisenberg and interlayer XX coupling can be engineered from native Ising interactions \cite{PhysRevA.109.L061304} through Floquet pulse sequences acting independently on the two layers \cite{Lukin_2020_Robust}. The Floquet toolbox underlying this construction is now well established across experimental platforms, including Rydberg arrays~\cite{Lukin_2020_Robust,Geier2021,Scholl2022}, polar molecules~\cite{Christakis2023,Miller2024}, and trapped ions~\cite{PRXQuantum.4.010334}.

In the unscaled model, with all couplings determined only by the power-law decay, $\lambda$ is fixed at unity. In this case, the ratio of interlayer to intralayer interaction strengths, and thereby the phase transition \cite{7g55-lpff}, is controlled by tuning the aspect ratio, $a_Z/L$. By introducing $\lambda$ as an independent control parameter, we now capture another physically distinct way to access the phase transitions.
Direct control of $\lambda$ can be realized via layer-selective Floquet pulses, which can rescale the interlayer XX couplings independently of the intralayer Heisenberg couplings at fixed geometry while preserving the spin structure of the interactions in Eq.~\eqref{eq:model}.

We initialize the dynamics from a state of oppositely polarized layers, $\langle\hat{S}_A\rangle = -\langle\hat{S}_B\rangle = (N/2)\,\hat{z}$, with $\hat{S}^\mu_\eta = \sum_{i\in\eta}\hat{s}^\mu_i$. This state is an eigenstate of the intralayer Heisenberg interactions, so all dynamics is driven by the interlayer XX exchange. We simulate the full nonequilibrium many-body spin dynamics using the discrete truncated Wigner approximation (dTWA) \cite{Schachenmayer2015,Zhu_NewJournalofPhysics_21_2019,Muleady_PRL_2023}, which has been benchmarked against tensor-network simulations for two-dimensional power-law spin models~\cite{Muleady_PRL_2023} and which we have validated against exact diagonalization for the present model~\cite{PhysRevA.109.L061304}.

\subsection{Bogoliubov Analysis}\label{sec:model_bog}
To gain an analytical understanding of  the resulting dynamics in the fully collective regime we map the spin system to bosonic excitations via a Holstein-Primakoff transformation of the layer spins~\cite{Holstein1940},
\begin{equation}
  \hat{S}^z_A = N/2 - \hat{a}^\dagger \hat{a},
  \qquad
  \hat{S}^+_A = \sqrt{N-\hat{a}^\dagger\hat{a}}\;\hat{a} \simeq \sqrt{N}\,\hat{a},
  \label{eq:HP}
\end{equation}
and an analogous mapping for layer $B$ with bosons $\hat{b}$ such that
$\hat{S}^z_B = -N/2 + \hat{b}^\dagger\hat{b}$ and $\hat{S}^-_B \simeq \sqrt{N}\,\hat{b}$, so that
the bosonic vacuum reproduces the initial state of oppositely polarized layers. At leading order in
$1/N$ the interlayer term reduces to a two-mode squeezing (TMS) form,
$\hat{H}_{\mathrm{TMS}} = (\lambda N V_{\mathrm{av}}/2)(\hat{a}^\dagger\hat{b}^\dagger + \hat{a}\hat{b})$,
with $V_{\mathrm{av}}$ the average interlayer interaction. The TMS results in squeezing of mixed quadratures $\hat{\mathcal{O}}^- = \hat{S}^x_A + \hat{S}^y_B$ (or $\hat{S}^y_A - \hat{S}^x_B$) and predicts an exponential increase/reduction of variances in time $\mathrm{Var}[\hat{\mathcal{O}}^{\pm}] =N/2 e^{\pm \lambda N V_{\mathrm{av}}t/\hbar}$.

To access the regime beyond the fully-collective approximation, we map individual spins to bosons $a_{\bm{j}}$ ($b_{\bm{j}}$) for layer A (B) as 
$ \hat{s}^+_{A \bm{j} } = \sqrt{1- \hat{a}_{\bm{j} } ^{\dagger} \hat{a}_{\bm{j} } }~ \hat{a}_{\bm{j} }$, $  \hat{s}^-_{A\bm{j} }    = \hat{a}^{\dagger}_{\bm{j} } \sqrt{1- \hat{a}_{\bm{j} } ^{\dagger} \hat{a}_{ \bm{j} } } $, $
      \hat{s}^z_{A\bm{j} }   = \frac{1}{2} - \hat{a}^{\dagger}_{\bm{j} } \hat{a}_{\bm{j} }   $, and expand to second order. 
The resulting quadratic Hamiltonian can then be diagonalised in momentum space via a Bogoliubov transformation~\cite{Bilitewski2023a,Bilitewski2023,7g55-lpff}. In the simplest case of a single atom in the unit cell, this yields quasi-energies
\begin{equation}
  E_{\bm{k}} \;=\; \sqrt{\varepsilon_{\bm{k}}^{\,2} - |\Omega_{\bm{k}}|^2},
  \label{eq:quasienergy}
\end{equation}
with $\varepsilon_{\bm{k}} = \epsilon_{\bm{k}} - \epsilon_{\bm{0}}, $ where $\epsilon_{\bm{k}}$ and $\Omega_{\bm{k}}$ are the Fourier transforms of the intralayer and interlayer exchange interactions, respectively. Modes with $|\Omega_{\bm{k}}| > |\varepsilon_{\bm{k}}|$ have imaginary eigenfrequencies and grow exponentially in time. As a consequence of the in-plane Heisenberg interaction in Eq.~\eqref{eq:model}, we have  $\varepsilon_{\bm{k}=\bm{0}} =0$, and the $\bm{k} = \bm{0}$ mode is always unstable (for non-vanishing average interlayer interaction $\Omega_0$), corresponding to the collective two-mode squeezing dynamics described above. Generically, additional finite-$\bm{k}$ modes may also be unstable depending on the model parameters $a_Z$, $L$, $\alpha$, and $\lambda$. %
We distinguish two regimes based on this momentum-space structure. A regime in which only the $\bm{k}=0$ mode is unstable, predicted to remain fully collective with all finite-$\bm{k}$ modes only weakly populated, and a regime with multiple unstable modes, in which finite-momentum excitations contribute to the dynamics. The boundary between these two regimes, defined by the condition $|\Omega_{\bm{k}}| = |\varepsilon_{\bm{k}}|$ for $\bm{k} \neq \bm{0}$, provides a quadratic-order prediction for the phase boundary.

\subsection{Phases and Metrological Utility}\label{sec:model_phases}
We identify the transition between the fully collective and partially collective phases within dTWA by using the system-size scaling of the minimum of the variance of the squeezed quadrature,
\begin{equation}
\mathrm{Var}[\hat{\mathcal{O}}^-]_{\min} =\min_t \mathrm{Var}[ \, \hat{\mathcal{O}}^-(t)]  \sim N^p,
\end{equation}
attained during the time evolution. The fully collective phase corresponds to a system-size-independent minimal variance ($p = 0$), while the partially collective phase is identified by the onset of system-size dependence, ($p > 0$). Physically, the two phases differ in how much of the dynamics remains confined to the collective $\bm{k}=\bm{0}$ manifold. In the fully collective phase the layer spins evolve as single macroscopic spins of essentially fixed length, finite-momentum modes remain weakly populated, and the minimal variance is set entirely by the collective TMS dynamics and is therefore independent of $N$. In the partially collective phase a finite fraction of the generated excitations is carried by finite-momentum modes, which reduces the collective spin length and terminates the squeezing earlier, so that the minimal variance acquires an explicit $N$ dependence. This distinction can be seen in the collective spin length and the layer polarization, which remain at their maximal values throughout the squeezing dynamics in the fully collective phase but are reduced in the partially collective phase~\cite{7g55-lpff}. We emphasize that the partially collective phase is not the non-collective limit. $p$ remains strictly below unity, so the squeezing is still scalable. Following standard arguments \cite{Sundar2023,PhysRevA.109.L061304}, the squeezed variance translates into an enhanced sensitivity for measuring a phase of rotation $\phi$ around $\hat{S}^z_A - \hat{S}^z_B$,
\begin{equation}
(\Delta\phi)^2 =   \frac{(\Delta\hat{\mathcal{O}^-})^2} {\langle \hat{S}^z_A - \hat{S}^z_B\rangle^2}.
\label{eq:phi}
\end{equation}
 Since the layer polarization remains of order $N$ at the time of optimal squeezing both in the fully collective and the partially collective phase \cite{7g55-lpff}, the sensitivity scales as $(\Delta\phi)^2 \sim N^{p-2}$. The fully collective phase $p=0$ therefore achieves the Heisenberg limit $(\Delta\phi)^2 \sim N^{-2}$, while the partially collective phase with $0<p<1$ retains scalable, quantum-enhanced sensitivity beyond the standard quantum limit $(\Delta\phi)^2_{\mathrm{SQL}} \sim N^{-1}$ of $N$ unentangled particles. Distinguishing the two phases via the
exponent $p$ is therefore equivalent to distinguishing two regimes of metrological utility.\\

The operators $\hat{\mathcal{O}}^{\pm} = \hat{S}^x_A \mp \hat{S}^y_B$ are mixed quadratures, each combining a transverse spin component of one layer with the orthogonal transverse component of the other, and in the bosonic language are the standard two-mode quadratures of the TMS state~\cite{PhysRevA.109.L061304}. Their physical role is fixed by the generator appearing in the sensitivity relation, Eq.~\eqref{eq:phi}, where $\phi$ is the phase of a rotation generated by $\hat{S}^{z}_{A} - \hat{S}^{z}_{B}$, that is, a rotation of equal magnitude and opposite direction about $z$ in the two layers. Such a staggered rotation is generated by any perturbation that couples with opposite sign to the two layers, e.g., a magnetic field. The bilayer therefore operates as a gradiometer and the squeezing generated here lowers the noise floor for measuring the field difference across the bilayer~\cite{RevMod_Metrology_2018}. Since the two layers are moreover spatially separated, the reduced variance of $\hat{\mathcal{O}}^{-}$ simultaneously certifies entanglement between two distinct ensembles, of the type used for EPR steering in spatially distributed atomic samples~\cite{Lange2018}.

\section{Phase-transitions across different lattice geometries, dimensions, and couplings\label{sec:phase}}
\begin{figure}[tb]
\includegraphics[width=\columnwidth]{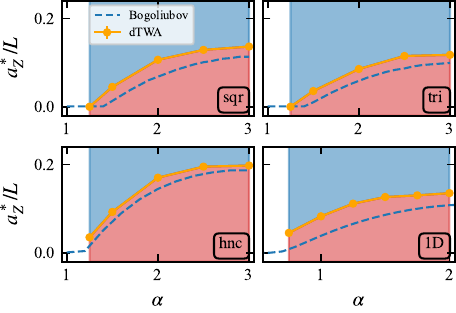}
\caption{Comparison of the phase-boundary within Bogoliubov analysis (dashed) and dTWA (solid with markers). Critical aspect ratio $a_Z^*/L$ as a function of power-law exponent $\alpha$ for $1\mathrm{D}$ and different $2\mathrm{D}$ lattice structures. Shading indicates the phases as determined by dTWA, fully collective (blue) and partially collective (pink). The dashed line is the Bogoliubov prediction, defined by the appearance of a second unstable mode at $\bm{k}\neq 0$, above which $\bm{k}= 0$ is the only unstable mode.
\label{fig:aZL_Vs_alpha}}
\end{figure} 
We first map out the dynamical phase boundary separating the fully collective and partially collective squeezing phases as a function of the lattice geometry and of the interlayer coupling strength, $\lambda$. This transition was established \cite{7g55-lpff} for square lattice geometry and one-dimensional ladders, and shown to be controlled by the aspect ratio $a_Z/L$ for a given range of power-law exponents $\alpha$. Here, we extend the analysis in two directions changing the microscopic interactions, while preserving their spin-structure. We consider $2\mathrm{D}$ bilayers built from triangular and honeycomb unit cells in addition to the square lattice, and we introduce an independent
control parameter $\lambda$ that rescales all interlayer couplings relative to the intralayer ones while preserving the spin structure and symmetries of the Hamiltonian of Eq.~\eqref{eq:model}.
 
The parameter $\lambda$ is motivated by experimental accessibility. In most platforms, the layer spacing $a_Z$ cannot be tuned freely over a wide range without simultaneously altering other microscopic parameters, whereas Floquet-engineered interactions provide direct control over the ratio of interlayer to intralayer couplings at fixed geometry. We show below that the dynamical transition can be driven purely by tuning this ratio, providing a practically accessible knob on the phase diagram.
 
We characterize the phase transitions using the two complementary approaches, Bogoliubov analysis and dTWA, described in Sec. \ref{sec:model}.
In the Bogoliubov analysis we use the aspect ratio $a_Z/L$ as the control parameter to locate the
boundary between the regime in which $\bm{k}=\bm{0}$ is the only unstable mode and the regime with
multiple unstable modes, which provides the quadratic-order prediction for the phase boundary. 
In parallel, we use dTWA to calculate the minimal variance, $\mathrm{Var}[\hat{\mathcal{O}}^-]_{\rm min}\sim N^{p}$, for identifying the transition from the fully collective phase with $p(a_Z/L>a_Z^*/L)=0$ to the partially collective phase with $p(a_Z/L\leq a_Z^*/L)>0$.

Within dTWA, at fixed lattice, $\alpha$, $\lambda$ and aspect ratio, we fit $\mathrm{Var}[\hat{\mathcal{O}}^-]_{\min} \propto N^{p}$ over the accessible system sizes, and assign the fully collective phase when the fitted $p$ is consistent with zero within its uncertainty and the partially collective phase otherwise. We identify $a^{*}_Z/L$ with the midpoint between the largest aspect ratio for which $p>0$ and the smallest for which $p=0$, taking half the spacing between these two values as its uncertainty. Appendix~\ref{app:identification} shows the underlying $p(a_Z/L)$ data and this extraction for representative cases. Within the Bogoliubov analysis the criterion is instead mode counting, with $a^{*}_Z$ defined by the value at which the largest unstable momentum drops to zero, as detailed in Appendix~\ref{app:bogoliubov}.

 \begin{figure}[t]
\includegraphics[width=\columnwidth]{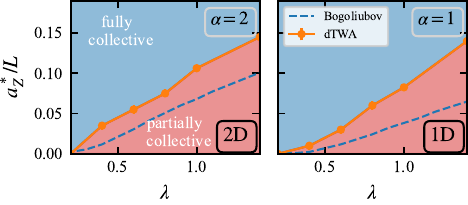}
\caption{
Comparison of the phase boundary within Bogoliubov analysis (dashed) and dTWA (solid with markers). Critical aspect ratio $a^{*}_Z/L$ as a function of the ratio $\lambda$ between interlayer and intralayer interactions, for $\alpha=2$ in 2D and $\alpha=1$ in 1D. Shading indicates the phases as determined by dTWA, fully collective (blue) and partially collective (pink). The dashed line is the Bogoliubov prediction, defined by the appearance of a second unstable mode at $\bm{k}\neq 0$, above which $\bm{k}= 0$ is the only unstable mode.
\label{fig:aZL_Vs_lambda}}
\end{figure} 
Fig.~\ref{fig:aZL_Vs_alpha} compares the two predictions across lattice geometries and dimensions at fixed $\lambda=1$. In all cases, the dTWA result (solid lines with markers) traces out the boundary between the fully collective phase (shaded blue) and partially collective phase (shaded pink). The Bogoliubov prediction (dashed lines) and the dTWA result show the same qualitative trend for all four cases. The critical aspect ratio $a_Z^*/L$ vanishes below a threshold value of $\alpha$ that depends on the lattice geometry and the dimensionality. Above this threshold value, $a_Z^*/L$ grows with increasing $\alpha$ within the range shown here. Further $\alpha$ values with short-range-like interaction are discussed in the next section. While generally we observe good qualitative agreement across all systems, a quantitative difference between Bogoliubov and dTWA is present for all geometries. The Bogoliubov analysis consistently underestimates the critical aspect ratio compared to dTWA. We attribute this systematic shift to non-linear corrections beyond the quadratic Bogoliubov theory, which destabilize the fully-collective phase; the Bogoliubov stability therefore provides a necessary, but not sufficient condition for collective dynamics.

Fig.~\ref{fig:aZL_Vs_lambda} shows the analogous comparison as a function of the interlayer coupling ratio $\lambda$ at fixed $\alpha$, in both $1\mathrm{D}$ and $2\mathrm{D}$. As $\lambda$ increases, the critical aspect ratio $a_Z^*/L$ grows monotonically: stronger interlayer coupling pushes the system further into the multiple-unstable-modes regime, and a larger geometric separation is required to recover fully collective behavior. For small $\lambda$, the transition occurs at very small aspect ratios/layer separations, making the fully collective phase accessible where it was prohibited for $\lambda=1$. Tuning $\lambda$ therefore provides a genuine control parameter to access the collective-partially collective transition. The relative strength of the interlayer interactions alone is sufficient to drive the system across the dynamical transition without any change in the underlying lattice geometry. We again observe good qualitative agreement of the Bogoliubov predictions and dTWA simulations.
 
Taken together, Figs.~\ref{fig:aZL_Vs_alpha} and~\ref{fig:aZL_Vs_lambda} demonstrate that the transition identified in Ref.~\cite{7g55-lpff} persists across a family of bilayer systems sharing the same (spin) symmetries, independent of lattice geometry, and is accessible via tuning the interlayer interactions with $\lambda$. The broad agreement between the Bogoliubov analysis and dTWA across lattice geometries, dimensions, and values of $\lambda$ further establishes the Bogoliubov approach as a reliable and computationally inexpensive predictor of the phase boundary for this class of models.

\section{Scaling of critical aspect ratio\label{sec:scaling_aZL}}

An important question is whether $a_Z/L$ is the correct control parameter of the transition across the full range of $\alpha$, i.e., whether $a_Z^*$ scales linearly with the system size $L$. We address this by combining a direct analysis of the Bogoliubov dispersion relations with dTWA data.
 
Within the Bogoliubov analysis, the $\bm{k}=\bm{0}$ mode (for our specific choice of spin interactions) is always unstable, and the transition to the partially collective phase occurs when a second mode, labeled $\bm{k}_1$, which turns out to be the smallest allowed momentum ($k_1=2\pi/L$ in $1\mathrm{D}$) becomes unstable. The critical condition is $\Omega_{\bm{k}_1} = \varepsilon_{\bm{k}_1}$, where $\Omega_{\bm{k}}$ is the Fourier transform of the interlayer interaction and $\varepsilon_{\bm{k}}$ the intralayer dispersion (see Sec.~\ref{sec:model_bog}). The numerical procedure used to locate $a^{*}_Z$ from this condition is given in Appendix~\ref{app:bogoliubov}. Moreover, we empirically find that for $\alpha >d+2$, close to the critical aspect ratio $a_Z^*/L$, $\Omega_{\bm{k}}$ shows only weak momentum dependence, allowing us to approximate $\Omega_{\bm{k}_1} \approx \Omega_0$.  
We may explain this in the following way: for interlayer pairs at separation $|\vect{r}_i - \vect{r}_j| = \sqrt{r^2+a_Z^2}$ with $r$ the in-plane distance, we have $\Omega_0 = \sum_r (r^2 + a_Z^2)^{-\alpha/2} \sim a_Z^{-\alpha} \int d^d r \, [1+(r/a_Z)^2]^{-\alpha/2} \propto a_Z^{d-\alpha}$ for $\alpha > d$, where we substituted $r^{\prime} = r/a_Z$. Thus, the interlayer interaction is integrable, and has a leading constant term with quadratic corrections negligible at $k_1 \sim 1/L$. Similarly, $\varepsilon_{\bm{k}}  \propto k^2 $ for $\alpha > d+2$, with a leading quadratic term. These analytical predictions are also confirmed in explicit numerical evaluations of the lattice Fourier transforms on finite systems (see Appendix~\ref{app:bogoliubov} and Figs.~\ref{fig:bogo_longrange}–\ref{fig:bogo_shortrange}). %

The smallest finite momentum accessible in a system of linear size $L$ is $k_1 \sim 1/L$, so that $\varepsilon_{\bm{k}_1}\propto 1/L^2$. Imposing $\Omega_0 = \varepsilon_{\bm{k}_1}$ yields $a_Z^{-\alpha+d}\, L^{2} = \mathrm{const}$ and therefore
\begin{equation}
  a_Z^* \;\propto\; L^{2/(\alpha-d)}
  \label{eq:aZ_scaling}
\end{equation}
for $\alpha > d+2$. Thus, the scaling exponent falls below unity, and $a_Z^*/L$ decreases with increasing system size $L$. %
For $\alpha = d+2$ logarithmic corrections enter the dispersion $\varepsilon_k \propto k^2 \ln(k)$\cite{PhysRevB.100.184306}, resulting in the prediction $a_Z^* \propto L/\sqrt{\ln(L)}$. %

We emphasize that this argument and the resulting scaling do not apply in the regime $\alpha < d+2$, where $\Omega_{\bm{k}}$ shows significant $k$-dependence, and the interplay of the momentum dependence of $\varepsilon_{\bm{k}}$ and $\Omega_{\bm{k}}$ results in the previously discussed scaling $a_Z^* \propto L$ (as also observed when numerically evaluating the exact condition $\Omega_{\bm{k}_1} = \varepsilon_{\bm{k}_1}$, see Appendix~\ref{app:bogoliubov}). %
\begin{figure}[t]
\includegraphics[width=\columnwidth]{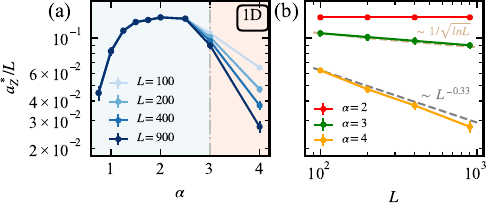}
\caption{System size dependence of critical aspect ratio $(a_Z^*/L)$ for different power-law exponents $\alpha$. (a) $a_Z^*/L$ as a function of $\alpha$ for different system size $L$. For $\alpha < d+2$ (light blue) $a_Z^*/L$ is independent of $L$, whereas for $\alpha \ge d+2$ it decreases with system size. (b) $a_Z^*/L$ as a function of $L$ for different $\alpha$. Dashed lines are the analytical predictions (see text).
\label{fig:large_alpha}}
\end{figure}
Figure~\ref{fig:large_alpha} tests these predictions against the dTWA data for the $1\mathrm{D}$ case, where large system sizes are accessible. Figure~\ref{fig:large_alpha}(a) shows $a_Z^*/L$ as a function of $\alpha$ for several values of $L$. For $\alpha < d+2$ (shaded blue region) the curves for different $L$ collapse, confirming that $a_Z^*/L$ is independent of system size and that $a_Z/L$ is the correct control parameter of the transition. For $\alpha > d+2$ the curves separate and $a_Z^*/L$ decreases systematically with $L$. %

Figure~\ref{fig:large_alpha}(b) plots $a_Z^*/L$ versus $L$ for three values of $\alpha$ representative of the distinct regimes. %
For $\alpha=2$, well within the long-range regime $\alpha<d+2$, the critical aspect ratio is independent of $L$. %
For $\alpha=3$, which sits exactly at $d+2$, the data shows weak decay confirming the prediction $a_Z^*/L\propto 1/\sqrt{\ln(L)}$. 
For $\alpha=4$, in the regime $\alpha > d+2$, we find a power-law decay with an exponent close to $-0.33$, in agreement with the analytical prediction $a_Z^*/L\propto L^{-1/3}$. %
 
The result has direct experimental implications. Dipolar interactions ($\alpha=3$) in $1\mathrm{D}$ sit exactly at the crossover, and in $2\mathrm{D}$ ($d=2$) lie within the long-range regime. In both cases $a_Z/L$ remains the correct control parameter at all accessible system sizes. For van der Waals Rydberg interactions ($\alpha=6$), Eq.~\eqref{eq:aZ_scaling} predicts access to the fully collective phase with smaller $a_Z^*/L$ for increasing system size.

\section{Universality across different geometries and coupling strengths\label{sec:universality}}
 \begin{figure}[t]\includegraphics[width=\columnwidth]{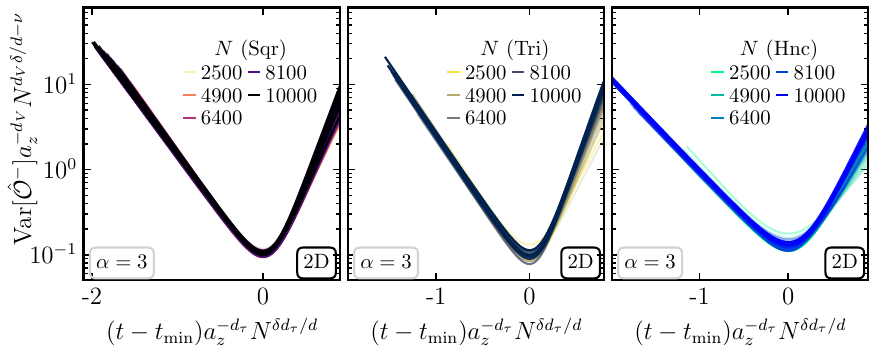}
\caption{Universality of the partially collective phase across different lattice structures. Rescaled variance ${\rm Var}[\hat{\mathcal{O}}^-] a_{Z}^{-d_{V}} N^{d_V\delta/d-\nu} $ vs rescaled time $(t - t_{\rm min})  a_{Z}^{-d_{\tau}} N^{\delta d_{\tau}/d}$ for different $a_Z$ values and system sizes $N$. For each $N$, a range of $a_Z$ values is plotted with increased fading for smaller values.\label{fig:Universality_lattice}
}
\end{figure}
We now turn to the critical scaling in the partially collective phase and examine the validity of the universality across two qualitatively different microscopic perturbations, each at a fixed value of power-law $\alpha$. 
We use the scaling ansatz proposed in Ref.~\cite{7g55-lpff} for the time evolution of the squeezed variance,
\begin{equation}
\mathrm{Var}\left[\hat{\mathcal{O}}^-\right]a_Z^{-d_{V}}N^{d_V\delta /d-\nu}=f\big [ (t-t_{\rm min}) a_Z^{-d_{\tau}} N^{\delta d_{\tau}/d} \big ]\label{eq:scaling_ansatz} 
\end{equation}
with four exponents $(d_V,d_\tau,\nu,\delta)$ characterizing the divergence of the variance and the relaxation timescale.

The form of Eq.~\eqref{eq:scaling_ansatz} is motivated by the observation that the variance minima occur at times $t_{\min}$ that grow systematically with $a_{Z}$, which suggests that the partially collective phase may obey a universal scaling that depends on a diverging relaxation timescale, in direct analogy with critical slowing down at an equilibrium critical point. The specific form of Eq.~\eqref{eq:scaling_ansatz} then follows from standard scaling theory \cite{Altland2010}, as set out in Ref.~\cite{7g55-lpff}. The layer spacing plays the role of the distance from criticality, and $d_{V}$ and $d_{\tau}$ describe respectively how the variance scale and the relaxation time diverge with $a_{Z}$ at fixed system size. The remaining system-size dependence is fixed by finite-size scaling. The distance from criticality must be measured on a system-size dependent scale, $a_{Z,\delta}=L^{-\delta}a_{Z}$, in terms of which the ansatz takes the compact form
\begin{equation}
\mathrm{Var}\!\left[\hat{\mathcal{O}}^-\right] a_{Z,\delta}^{-d_{V}} N^{-\nu}
= f\!\big[(t-t_{\rm min})\,a_{Z,\delta}^{-d_{\tau}}\big],
\end{equation}
with $\nu$ setting the residual $N$ dependence of the variance. We stress that $\delta$ is an independent exponent and is not fixed by the location of the phase boundary. The transition occurs at $a_Z^{*}\propto L$ for $\alpha<d+2$, but the scale that collapses the dynamics within the partially collective phase is in general a different one, and we find $\delta$ to vary systematically with $\alpha$ and $d$. The four exponents are not independent. Setting $t = t_{\min}$ in the ansatz yields the scaling exponent $p = d_{V}(1-\delta)/d + \nu$, which acts as the order parameter for this transition and which we use as a constraint when extracting $\nu$.
We now apply the scaling ansatz of Eq.~\eqref{eq:scaling_ansatz} to a range of lattices for various values of $\alpha$ and across the microscopic control parameter, $\lambda$. 
 
Fig.~\ref{fig:Universality_lattice} shows the scaling collapse for square, triangular, and honeycomb $2\mathrm{D}$ bilayers with $\alpha=3$, the physically relevant case corresponding to dipolar interactions. For each geometry we plot the rescaled variance against rescaled time for a range of system sizes $N$ and of aspect ratios $a_Z/L$ within the partially collective phase. %
The data collapse onto a single scaling function in each panel, and the exponents extracted for the three geometries, listed in the upper block of Table~\ref{tab:main_exponents}, agree within error. Appendix~C further extends the lattice-geometry comparison to $\alpha = 1.5$, $2.0$ and $2.5$, with critical exponents at each $\alpha$ consistent across the three lattices within error.
 
\begin{figure}[t]
\includegraphics[width=\columnwidth]{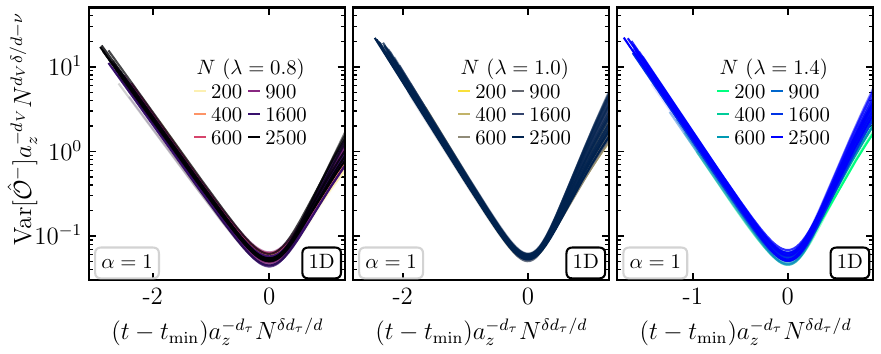}
\caption{Universality of the partially collective phase across different interlayer coupling strengths $\lambda$. Rescaled variance ${\rm Var}[\hat{\mathcal{O}}^-] a_{Z}^{-d_{V}} N^{d_V\delta/d-\nu} $ vs rescaled time $(t - t_{\rm min})  a_{Z}^{-d_{\tau}} N^{\delta d_{\tau}/d}$ for different $a_Z$ values and system sizes $N$. For each $N$, a range of $a_Z$ values is plotted with increased fading for smaller values.
\label{fig:Universality_lambdas}
}
\end{figure}

Figure~\ref{fig:Universality_lambdas} presents the case where the lattice is fixed ($1\mathrm{D}$ ladder, $\alpha=1$) and the interlayer coupling ratio $\lambda$ is varied across three values $\lambda=0.8,\,1.0,\,1.4$ that span either side of the unscaled model with $\lambda=1$. While leaving the (spin) symmetry structure of the Hamiltonian invariant, $\lambda$ rescales interlayer coupling, shifts the phase boundary (Fig.~\ref{fig:aZL_Vs_lambda}), and modifies the Bogoliubov spectrum of unstable modes. The scaling collapse is nonetheless equally effective for all three values of $\lambda$, and the extracted exponents, given in the lower block of Table~\ref{tab:main_exponents}, are again consistent within error.\\

The scaling exponents can be related to physical parameters of the model. We found that in the partially collective phase the minimal variance scales as $N^p$, which for two-mode squeezing is directly related to the number of generated pairs as $\mathrm{Var}_{\mathrm{min}} \sim N/N_{\mathrm{pair}}$.
Therefore, at the point of minimal variance $e^{\Gamma_0 t_{\mathrm{min}}} = N^{1-p}$ and consequently $t_{\mathrm{min}} = \ln(N) (1-p) /\Gamma_0$, where  $\Gamma_{0} \propto |\Omega_0|$ is the collective growth rate. For the parameters of Fig.~\ref{fig:Universality_lattice}, $\alpha = 3$ in 2D, the Bogoliubov excitation spectrum is exactly solvable and $\Gamma_{0} \propto 1/a_{Z}$, resulting in $d_{\tau} = 1$. Our extracted values, $d_{\tau} = 1.12$, $1.05$ and $1.07$ for the square, triangular and honeycomb lattices respectively, are consistent with this prediction in all three cases.\\
The exponent $\delta$ admits a similarly direct interpretation. Reference~\cite{7g55-lpff} argued on physical grounds that $\delta > 1$ for genuinely long-range interactions, $\alpha < d$, since this is what causes the accessible window of the partially collective phase to shrink with increasing system size. Indeed, for the 2D square lattice we obtain $\delta = 4.20\pm 0.30$ at $\alpha = 1.5 < d$, $\delta = 1.06\pm 0.16$ at $\alpha = 2 = d$, and for $\alpha > d$ we find $\delta = 0.34 \pm 0.08$ ($\alpha = 2.5$) and $\delta = 0.15 \pm 0.06$ ($\alpha = 3$). This same trend is found in all lattices studied and in 1D. This suggests that indeed the exponent $\delta$ crosses unity at the boundary between genuinely long-range interactions and integrable power laws. The remaining two exponents are tied more directly to the observable variance and squeezing. We find $d_{V} < 0$ in every case, so that at fixed $N$ the minimal variance decreases with increasing layer spacing and approaches the transition from within the partially collective phase with improving achievable squeezing. The exponent $p$ carries a direct metrological meaning, since $(\Delta\phi)^{2} \sim N^{p-2}$, so that $p = 0$ corresponds to the Heisenberg limit and $0 < p < 1$ to scalable, quantum-enhanced sensitivity beyond the standard quantum limit. The exponent $p$ therefore distinguishes between the two regimes of metrological utility discussed in Sec.~\ref{sec:model_phases}. We note finally that the broader picture emerging from these observations is that while the exponents are insensitive to lattice geometry and to the coupling ratio $\lambda$, they do depend on $\alpha$ and $d$.

The two tests probe universality in qualitatively different, complementary ways: varying lattice geometry changes the structure of the couplings, and hence the behavior of $\varepsilon_{\bm{k}}$ at finite $k$, while varying $\lambda$ uniformly rescales the pairing strength $\Omega_{\bm{k}}$ relative to $\varepsilon_{\bm{k}}$. The agreement of the critical exponents under both types of perturbation therefore provides substantially stronger evidence that they are insensitive to microscopic detail and depend only on the symmetries of the model, defining a genuine nonequilibrium universality class.
 
\begin{table}[t]
\caption{Critical exponents in the partially collective phase for the parameters of Figs.~4 and 5: $\alpha=3$ in 2D for the square, triangular and honeycomb lattices (upper block), and $\alpha=1$ in 1D for three values of the interlayer coupling ratio $\lambda$ (lower block). Uncertainties on the final digits are given in parentheses. The exponent $\nu$ follows from the scaling relation $\nu = p - d_V(1-\delta)/d$. Values for all $\alpha$ studied are given in Appendix~C.}
\label{tab:main_exponents}
\begin{ruledtabular}
\begin{tabular}{lccccc}
 & $d_V$ & $d_\tau$ & $\delta$ & $\nu$ & $p$ \\
\hline
Square & $-1.29(22)$ & $1.12(8)$ & $0.15(6)$ & $0.82(38)$ & $0.275(65)$ \\
Triangular & $-1.18(22)$ & $1.05(10)$ & $0.16(8)$ & $0.68(15)$ & $0.185(110)$ \\
Honeycomb & $-1.14(28)$ & $1.07(8)$ & $0.11(9)$ & $0.78(17)$ & $0.270(100)$ \\
\hline
$\lambda=0.8$ & $-0.60(16)$ & $0.24(6)$ & $1.05(7)$ & $0.21(7)$ & $0.24(6)$ \\
$\lambda=1.0$ & $-0.61(16)$ & $0.25(5)$ & $1.07(12)$ & $0.21(8)$ & $0.25(6)$ \\
$\lambda=1.4$ & $-0.62(14)$ & $0.24(6)$ & $1.02(7)$ & $0.29(7)$ & $0.30(6)$ \\
\end{tabular}
\end{ruledtabular}
\end{table}

\section{Outlook\label{sec:outlook}}
In this work we have extended the framework of dynamical squeezing phase transitions established in \cite{7g55-lpff} along several directions. We have shown that the dynamical transitions between fully collective and partially collective squeezing persist across square, triangular, and honeycomb $2\mathrm{D}$ bilayers and $1\mathrm{D}$ ladders for different power-law exponents $\alpha$. We further show that the transition can be driven at fixed geometry purely through interaction engineering via the interlayer coupling ratio $\lambda$. The qualitative agreement between the Bogoliubov analysis and dTWA simulations across all these settings establishes the Bogoliubov instability criterion as a reliable, computationally inexpensive predictor of the phase boundary. Building on this, we derived an analytical scaling of the transition $a_Z^* \propto L^{2/(\alpha-d)}$ for $\alpha > d+2$, in agreement with our dTWA data. This reveals a previously unrecognized change of scaling behavior for short-range-dominated interactions where $a_Z/L$ ceases to be the correct scaling variable. The universal scaling of the variance in the partially collective phase was found to be preserved both under changes of lattice geometry at the experimentally relevant value $\alpha = 3$ and under rescalings of the interlayer coupling strength, with critical exponents consistent within error across all cases. Together, these results provide substantially broader evidence for a genuine nonequilibrium universality class than was previously available, and place this universality on firmer analytical footing.

These findings have direct experimental consequences. The ability to drive the transition by tuning $\lambda$ alone removes the requirement of physically rearranging the bilayer geometry, which is typically challenging in experimental platforms. Floquet-engineered interactions of the type required have been demonstrated across a wide range of architectures, including Rydberg arrays \cite{Lukin_2020_Robust,Geier2021,Scholl2022}, polar molecules \cite{Christakis2023,Miller2024}, and trapped ions \cite{PRXQuantum.4.010334}, and combine naturally with bilayer geometries already realized in polar molecules \cite{Tobias2022}, magnetic atoms \cite{Du2024}, Rydberg tweezer arrays \cite{PhysRevLett.130.180601,Barredo2018,Bluvstein2023}, and trapped ions \cite{Hawaldar2024}.

The universality identified in our work is tied to the combination of intralayer SU(2) (Heisenberg) and interlayer U(1) (XX) interactions of the Hamiltonian \cite{7g55-lpff}, and the present results should accordingly be understood as establishing universality within a symmetry class. The Floquet-engineering toolbox that enables tuning $\lambda$ in the symmetric case can equally be used to introduce controlled symmetry-breaking perturbations. Clarifying how the critical exponents evolve as these symmetries are broken, either by anisotropic interactions or by external fields, would delineate the boundaries of the universality class and connect to broader questions about the role of symmetry in classifying nonequilibrium phases of matter.  A natural further direction is to clarify the relationship between the exponents $(d_V,d_\tau,\nu,\delta)$ and those identified in related studies of non-thermal fixed points in short-range Heisenberg models \cite{Rodriguez_Nieva_2022}.

\begin{acknowledgments}
\noindent{\textit{Acknowledgements:}}
This material is based upon work supported by the Air Force Office of Scientific Research under Award No. FA9550-25-1-0340. TB acknowledges support from the National Science Foundation through NRT NSF Grant No. DGE-2510202. The computing for this project was performed at the High Performance Computing Center at Oklahoma State University supported in part through the National Science Foundation grant OAC-1531128. 
\end{acknowledgments}

\section*{Data Availability}
The data that support the findings of this article are publicly available at \cite{Duha2026-rb}.

\bibliography{main}{}


\appendix
\section{Bogoliubov Analysis}\label{app:bogoliubov}

In this appendix we provide additional details on the Bogoliubov analysis introduced in Sec.~\ref{sec:model} and used in Sec.~\ref{sec:scaling_aZL} to derive the analytical scaling of the critical aspect ratio with system size. Within the quadratic Bogoliubov theory, the dynamics in momentum space is governed by the intralayer dispersion $\varepsilon_{\bm{k}}$ and the Fourier transform of the interlayer interaction $\Omega_{\bm{k}}$, with quasi-energies $E_{\bm{k}} = \sqrt{\varepsilon_{\bm{k}}^2 - |\Omega_{\bm{k}}|^2}$. The $\bm{k}=\bm{0}$ mode is always unstable, and the transition out of the fully collective phase occurs when the smallest finite momentum of order $k_1 \sim 1/L$ becomes unstable, i.e. when $|\Omega_{\bm{k}_1}| = \varepsilon_{\bm{k}_1}$. The two regimes $\alpha < d+2$ and $\alpha > d+2$ are distinguished by the long-wavelength behavior of $\varepsilon_{\bm{k}}$ and $\Omega_{\bm{k}}$, and lead to qualitatively different scaling of the critical aspect ratio $a_Z^*/L$ with $L$.\\

\begin{figure}
\includegraphics[width=\columnwidth]{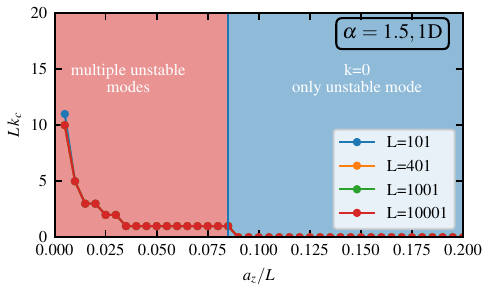}
\caption{Largest unstable mode $k_c$ scaled by linear system size $L$, as a function of the aspect ratio $a_Z/L$ for a range of system sizes $L$ (legend) in 1D for $\alpha=1.5$. 
\label{fig:fig_Bog_1}}
\end{figure}

We first specify how $a^{*}_Z$ is extracted. For a given lattice, dimension $d$, power law $\alpha$, layer spacing $a_Z$, coupling ratio $\lambda$ and linear size $L$, we evaluate the Fourier transforms $\varepsilon_{\bm k}$ and $\Omega_{\bm k}$ directly on the discrete momenta allowed by the finite system, using the same real-space interaction matrix $V_{ij}$ as in the dTWA simulations. A mode is unstable when $ \varepsilon^2_{\bm k} < |\Omega_{\bm k}|^2 $, and since $\varepsilon_0 = 0$ the $\bm{k}=0$ mode is unstable for any non-vanishing $\Omega_0$. We then determine the largest momentum $k_c$ with imaginary Bogoliubov eigenenergy as a function of $a_Z$, and identify $a^{*}_Z$ with the point at which $k_c$ drops discontinuously to zero, so that the number of unstable modes changes from many to one. This diagnostic is shown explicitly in Fig.~\ref{fig:fig_Bog_1}, where $Lk_c$ is plotted against $a_Z/L$ and the transition point is seen to be sharp and, in the long-range regime, independent of system size. The same criterion is used throughout, both for the dashed curves in Figs.~\ref{fig:aZL_Vs_alpha} and \ref{fig:aZL_Vs_lambda} and for the data in Figs.~\ref{fig:bogo_longrange} and \ref{fig:bogo_shortrange}. In the short-range regime $\alpha > d+2$ it continues to apply, but the resulting $a^{*}_Z$ no longer scales linearly with $L$, which is the effect reported in Sec.~\ref{sec:scaling_aZL}.
 
\subsection{Long-range regime: \texorpdfstring{$\alpha < d+2$}{alpha < d+2}}

\begin{figure}
\includegraphics[width=\columnwidth]{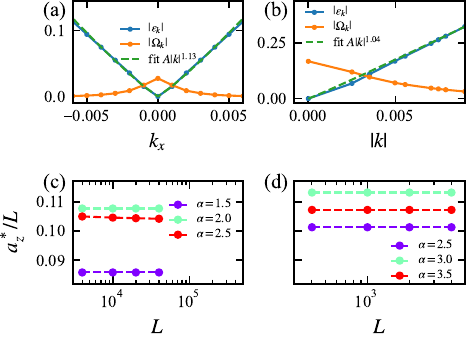}
\caption{Bogoliubov analysis in the long-range regime ($\alpha < d+2$). (a),(b) Long-wavelength behavior of the intralayer dispersion $\varepsilon_{\bm{k}}$ and the interlayer Fourier transform $|\Omega_{\bm{k}}|$ for (a) $\alpha = 2$ in $1\mathrm{D}$ and (b) $\alpha = 3$ in $2\mathrm{D}$. Dashed lines show fits of $\varepsilon_{\bm{k}}$ to $A|k|^s$, yielding sub-quadratic exponents $s < 2$, consistent with $s=\alpha -d$, while $|\Omega_{\bm{k}}|$ shows significant momentum dependence near $k=0$. (c),(d) Critical aspect ratio $a_Z^*/L$ as a function of linear system size $L$ for several values of $\alpha$ in the long-range regime in (c) $1\mathrm{D}$ and (d) $2\mathrm{D}$.\label{fig:bogo_longrange}
}
\end{figure}

Fig.~\ref{fig:bogo_longrange} shows the long-wavelength behavior of $\varepsilon_{\bm{k}}$ and $\Omega_{\bm{k}}$ for representative values of $\alpha = 2$ for $1\mathrm{D}$ systems in Fig.~\ref{fig:bogo_longrange}(a) and $\alpha = 3$ for $2\mathrm{D}$ systems in Fig.~\ref{fig:bogo_longrange}(b), evaluated numerically from the Fourier transforms on finite systems. In contrast to the short-range case discussed below, $\Omega_{\bm{k}}$ exhibits a significant momentum dependence at small $k$. The intralayer dispersion $\varepsilon_{\bm{k}}$ likewise deviates from a simple quadratic form, with a fit to $|k|^s$ giving exponents $s < 2$ that depend on $\alpha$. The instability condition $|\Omega_{\bm{k}_1}| = \varepsilon_{\bm{k}_1}$ must therefore be evaluated using the full $k$-dependence of both quantities, and the simple approximation $\Omega_{\bm{k}_1} \approx \Omega_0$ used in the short-range case (see below) does not apply.
 
The bottom panels of Fig.~\ref{fig:bogo_longrange} show the resulting critical aspect ratio $a_Z^*/L$ as a function of $L$ for several values of $\alpha$ in the long-range regime for $1\mathrm{D}$ (Fig.~\ref{fig:bogo_longrange}(c)) and for $2\mathrm{D}$ (Fig.~\ref{fig:bogo_longrange}(d)) systems, obtained by numerically solving $|\Omega_{\bm{k}_1}| = \varepsilon_{\bm{k}_1}$. For all $\alpha$ considered, the critical aspect ratio is essentially independent of $L$ over the accessible range of system sizes, confirming that $a_Z^* \propto L$ and that $a_Z/L$ remains the correct control parameter of the transition. This recovers the empirical observation of Ref.~\cite{7g55-lpff} and is consistent with the dTWA data in Fig.~\ref{fig:large_alpha}(a) of the main text in the shaded $\alpha < d+2$ region.

\begin{figure}[t]\includegraphics[width=\columnwidth]{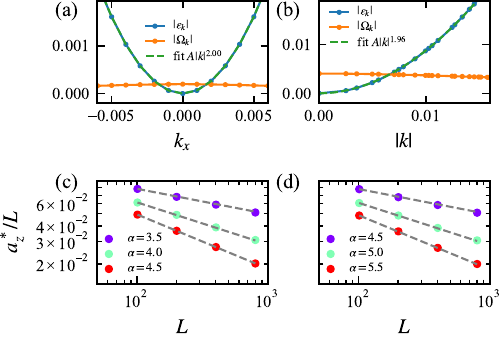}
\caption{Bogoliubov analysis in the short-range regime ($\alpha > d+2$). (a),(b) Long-wavelength behavior of the intralayer dispersion $\varepsilon_{\bm{k}}$ and the interlayer Fourier transform $|\Omega_{\bm{k}}|$ for (a) $\alpha = 4$ in $1\mathrm{D}$ and (b) $\alpha = 5$ in $2\mathrm{D}$. Dashed lines show fits of $\varepsilon_{\bm{k}}$ to $A|k|^s$, yielding $s \approx 2$, while $|\Omega_{\bm{k}}|$ is essentially flat at small $k$, justifying the approximation $\Omega_{\bm{k}_1} \approx \Omega_0$. (c),(d) Critical aspect ratio $a_Z^*/L$ as a function of linear system size $L$ for several values of $\alpha$ in the short-range regime in (c) $1\mathrm{D}$ and (d) $2\mathrm{D}$. Dashed grey lines are analytical predictions $a_Z^* \propto L^{2/(\alpha-d)}$.\label{fig:bogo_shortrange}
}
\end{figure}

\subsection{Short-range regime: \texorpdfstring{$\alpha > d+2$}{alpha > d+2}}
 
Figure~\ref{fig:bogo_shortrange} shows the analogous results for $\alpha$ in the short-range regime. The top panels demonstrate that $\Omega_{\bm{k}}$ becomes essentially flat at small $k$, justifying the approximation $\Omega_{\bm{k}_1} \approx \Omega_0$ used in the derivation of Eq.~\eqref{eq:aZ_scaling}. The intralayer dispersion $\varepsilon_{\bm{k}}$ recovers a clean quadratic form $\varepsilon_{\bm{k}} \propto |k|^2$, as expected, with the fitted exponents $s \approx 2$ across the values of $\alpha$ considered. %
With $\Omega_0 \propto a_Z^{-\alpha+d}$ and $\varepsilon_{\bm{k}_1}\propto k_1^2 \propto 1/L^2$, imposing $\Omega_0 = \varepsilon_{\bm{k}_1}$ yields $a_Z^* \propto L^{2/(\alpha-d)}$, Eq.~\eqref{eq:aZ_scaling} of the main text.
 
The bottom panels of Fig.~\ref{fig:bogo_shortrange} show the critical aspect ratio as a function of $L$, for several values of $\alpha$ in the
short-range regime. In contrast to the long-range case, $a_Z^*/L$ now decreases with $L$, with the predicted power-law dependence $L^{2/(\alpha-d) - 1}$ providing a good description of the data. This confirms that $a_Z/L$ ceases to be the correct control parameter in the
short-range regime and matches the dTWA result in Fig.~\ref{fig:large_alpha}(b) of the main text, in particular the $L^{-0.33}$ scaling observed at $\alpha=4$ in $1\mathrm{D}$.

\section{Identification of Phase Transitions from dTWA data}\label{app:identification}

\begin{figure}[t]\includegraphics[width=\columnwidth]{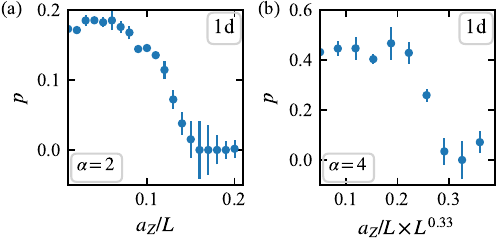}
\caption{Minimal variance system size scaling exponent $p$ versus $a_Z/L$ for (a) $\alpha=2$ and (b) $\alpha=4$ cases,  where $\text{Var}[\hat{\mathcal{O}}^-]_{\rm min} \sim N^{p}$. The scaling exponent $p$ jumps from zero to a nonzero value at the critical $a_Z^*/L$.}
\label{fig:p_vs_aZL}
\end{figure}

In Sec.~\ref{sec:phase} we identify the dTWA phase boundary by extracting the critical aspect ratio $a_Z^*/L$ at which the system-size scaling exponent $p$ of the minimal variance, $\mathrm{Var}[\hat{\mathcal{O}}^-]_{\min} \sim N^p$, departs from zero. Figure~\ref{fig:p_vs_aZL} illustrates this procedure for two representative cases, $\alpha=2$ and $\alpha=4$ in $1\mathrm{D}$, corresponding to the long-range and short-range regimes discussed in Appendix~\ref{app:bogoliubov}.
 
In each panel the exponent $p$ is obtained from a fit of the minimal variance to a power law in $N$ at fixed aspect ratio, using the set of
system sizes accessible in dTWA. In the fully collective phase $p$ is consistent with zero within error, while in the partially collective phase
$p$ takes a finite positive value. The critical aspect ratio is identified with the position at which $p$ transitions between these two behaviors. For $\alpha=2$ [Fig.~\ref{fig:p_vs_aZL}(a)] the horizontal axis is $a_Z/L$, and the transition occurs at a system-size-independent value, in agreement with the Bogoliubov prediction $a_Z^* \propto L$. For $\alpha=4$ [Fig.~\ref{fig:p_vs_aZL}(b)] the horizontal axis is rescaled as $(a_Z/L)\,L^{0.33}$, which collapses the data for different $N$ onto a common transition point, consistent with the predicted scaling
$a_Z^* \propto L^{2/3}$, i.e. $a_Z^*/L \propto L^{-1/3}$. The transition points extracted in this way provide the markers in Figs.~\ref{fig:aZL_Vs_alpha} and \ref{fig:aZL_Vs_lambda} of the main text.

\section{Determination of critical exponents \label{app:critical_exponents}}

\begin{table*}
\begin{ruledtabular}
\begin{tabular}{lcccccccccccccc}
Lattice & $\alpha$ & $d_V$ & $\Delta d_V$ & $d_\tau$ & $\Delta d_\tau$ & $\mathcal{S}_{\min}^{(d_V,d_\tau)}$ & $\delta$ & $\Delta\delta$ & $\mathcal{S}_{\min}^{(\delta)}$ & $\nu$ & $\Delta\nu$ & $p$ & $\Delta p$ & $\mathcal{S}_{\min}^{(p)}$ \\
\hline
Sqr & $1.5$ & $-0.88$ & $0.34$ & $0.16$ & $0.08$ & $1.14$ & $4.20$ & $0.30$ & $6.00$ & $-1.37$ & $0.70$ & $0.035$ & $0.055$ & $0.50$ \\
Sqr & $2.0$ & $-1.11$ & $0.22$ & $0.38$ & $0.06$ & $2.24$ & $1.06$ & $0.16$ & $2.39$ & $0.14$ & $0.07$ & $0.175$ & $0.055$ & $0.63$ \\
Sqr & $2.5$ & $-1.22$ & $0.22$ & $0.70$ & $0.08$ & $1.27$ & $0.34$ & $0.08$ & $1.19$ & $0.65$ & $0.23$ & $0.245$ & $0.065$ & $0.48$ \\
Sqr & $3.0$ & $-1.29$ & $0.22$ & $1.12$ & $0.08$ & $2.20$ & $0.15$ & $0.06$ & $2.14$ & $0.82$ & $0.38$ & $0.275$ & $0.065$ & $0.61$ \\
\hline
Tri & $1.5$ & $-0.70$ & $0.21$ & $0.10$ & $0.06$ & $2.75$ & $6.40$ & $0.60$ & $2.46$ & $-1.86$ & $0.61$ & $0.030$ & $0.085$ & $2.50$ \\
Tri & $2.0$ & $-1.11$ & $0.24$ & $0.38$ & $0.07$ & $2.22$ & $1.04$ & $0.16$ & $3.28$ & $0.15$ & $0.07$ & $0.175$ & $0.060$ & $0.53$ \\
Tri & $2.5$ & $-1.16$ & $0.16$ & $0.67$ & $0.08$ & $1.47$ & $0.37$ & $0.12$ & $3.00$ & $0.60$ & $0.13$ & $0.235$ & $0.100$ & $2.24$ \\
Tri & $3.0$ & $-1.18$ & $0.22$ & $1.05$ & $0.10$ & $2.47$ & $0.16$ & $0.08$ & $4.05$ & $0.68$ & $0.15$ & $0.185$ & $0.110$ & $3.60$ \\
\hline
Hnc & $1.5$ & $-0.80$ & $0.28$ & $0.14$ & $0.08$ & $1.15$ & $4.78$ & $0.42$ & $2.75$ & $-1.38$ & $0.56$ & $0.130$ & $0.100$ & $1.66$ \\
Hnc & $2.0$ & $-0.95$ & $0.18$ & $0.35$ & $0.06$ & $1.48$ & $1.06$ & $0.16$ & $1.38$ & $0.19$ & $0.07$ & $0.220$ & $0.055$ & $0.46$ \\
Hnc & $2.5$ & $-1.20$ & $0.28$ & $0.64$ & $0.08$ & $1.01$ & $0.32$ & $0.13$ & $1.71$ & $0.66$ & $0.16$ & $0.255$ & $0.100$ & $1.89$ \\
Hnc & $3.0$ & $-1.14$ & $0.28$ & $1.07$ & $0.08$ & $0.72$ & $0.11$ & $0.09$ & $3.96$ & $0.78$ & $0.17$ & $0.270$ & $0.100$ & $1.68$ \\

\end{tabular}
\end{ruledtabular}
\caption{Critical exponents extracted from the scaling collapse for the partially collective phase in 2D, for square (Sqr), triangular (Tri), and honeycomb (Hnc) lattice geometries across a range of power-law exponents $\alpha$. For each exponent we list the value, its uncertainty, and the corresponding minimum cost function $\mathcal{S}_{\min}$ from the optimization procedure described in the text.}\label{tab:exponents_geom}
\end{table*}
 
\begin{table*}
\begin{ruledtabular}
\begin{tabular}{ccccccccccccccc}
$\lambda$ & $\alpha$ & $d_V$ & $\Delta d_V$ & $d_\tau$ & $\Delta d_\tau$ & $\mathcal{S}_{\min}^{(d_V,d_\tau)}$ & $\delta$ & $\Delta\delta$ & $\mathcal{S}_{\min}^{(\delta)}$ & $\nu$ & $\Delta\nu$ & $p$ & $\Delta p$ & $\mathcal{S}_{\min}^{(p)}$ \\
\hline

$0.8$ & $1.0$ & $-0.60$ & $0.16$ & $0.24$ & $0.06$ & $4.70$ & $1.05$ & $0.07$ & $4.38$ & $0.21$ & $0.07$ & $0.24$ & $0.06$ & $1.83$ \\
$1.0$ & $1.0$ & $-0.61$ & $0.16$ & $0.25$ & $0.05$ & $3.31$ & $1.07$ & $0.12$ & $3.04$ & $0.21$ & $0.08$ & $0.25$ & $0.06$ & $1.30$ \\

$1.4$ & $1.0$ & $-0.62$ & $0.14$ & $0.24$ & $0.06$ & $2.72$ & $1.02$ & $0.07$ & $4.50$ & $0.29$ & $0.07$ & $0.30$ & $0.06$ & $1.75$ \\
\end{tabular}
\end{ruledtabular}
\caption{Critical exponents extracted from the scaling collapse for the partially collective phase in 1D for $\lambda = 0.8$, $\lambda = 1.0$ and $\lambda = 1.4$. For each exponent we list the value, its uncertainty, and the corresponding minimum cost function $\mathcal{S}_{\min}$ from the optimization procedure described in the text.}\label{tab:exponents_lambda}
\end{table*}

 \begin{figure*}[hbtp]
\begin{center}

\subfloat[Square, $N = 10000$]{
    \includegraphics[width = 7cm]{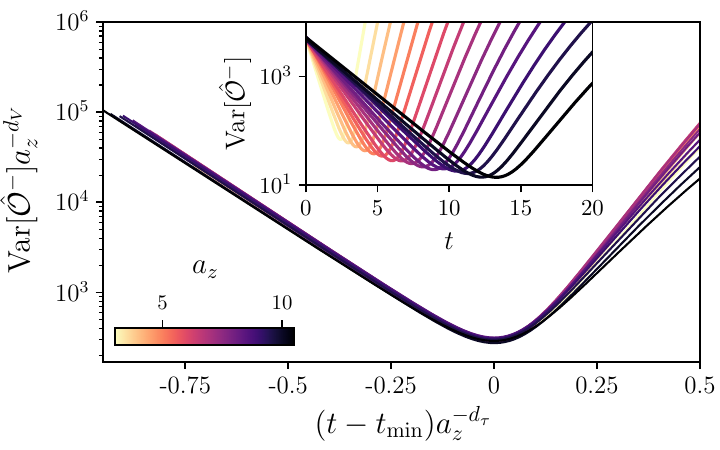} }
\subfloat[Square]{
    \includegraphics[width = 7cm]{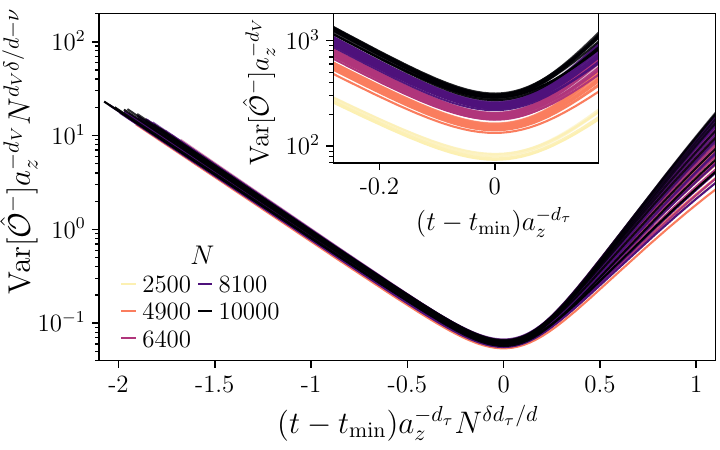} }

\subfloat[Triangular, $N = 10000$]{
    \includegraphics[width = 7cm]{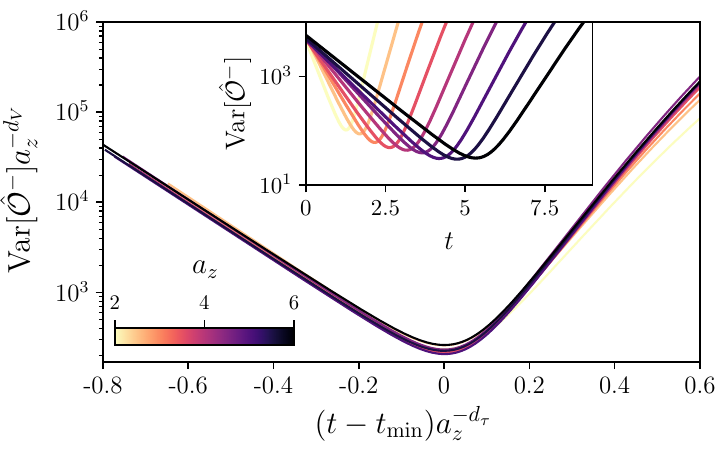} }
\subfloat[Triangular]{
    \includegraphics[width = 7cm]{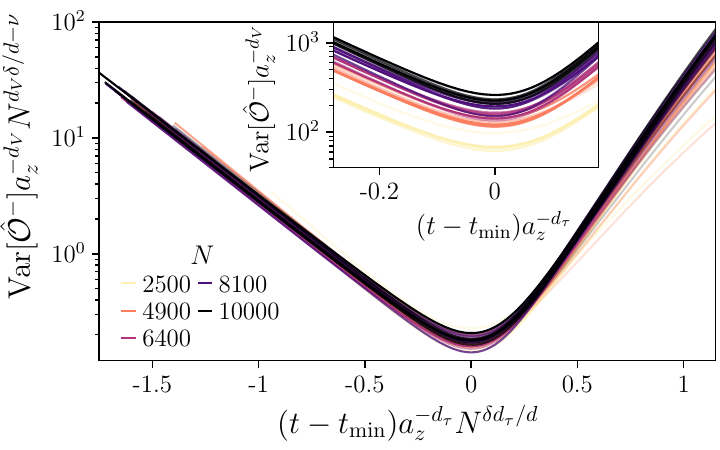} }

\subfloat[Honeycomb, $N = 10000$]{
    \includegraphics[width = 7cm]{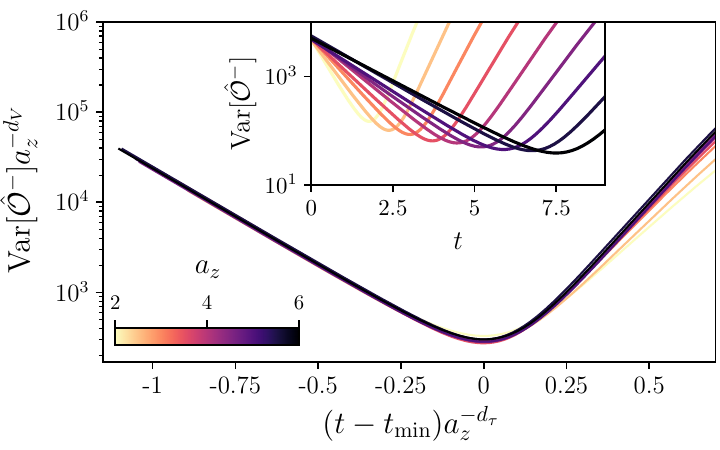} }
\subfloat[Honeycomb]{
    \includegraphics[width = 7cm]{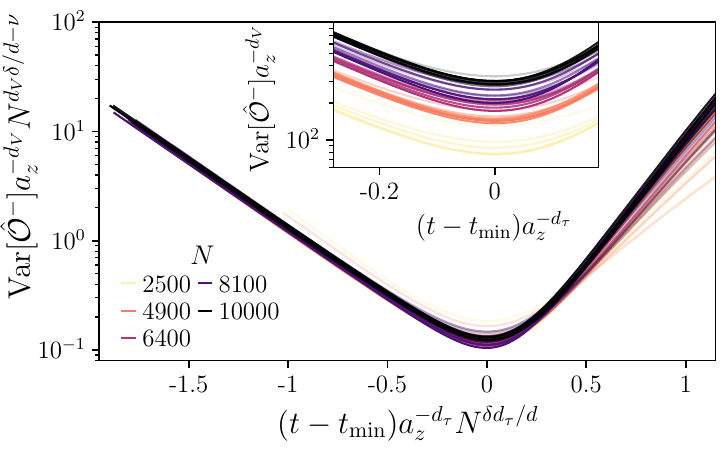} }

\end{center}
    \caption{Raw data and scaling collapses for $\alpha=3$ in $2\mathrm{D}$, for the square lattice  [(a) and (b)], triangular lattice  [(c) and (d)] and the honeycomb lattice  [(e) and (f)]. Lines for smaller $a_Z$ values are more faded in (b), (d) and (f). The exponents used for the collapse are given in Table~\ref{tab:exponents_geom}.  \label{fig:collapse_geom}
    }
\end{figure*}

\begin{figure*}[hbtp]
\begin{center}
\subfloat[$\lambda = 0.8$, $N = 2500$]{
    \includegraphics[width = 7cm]{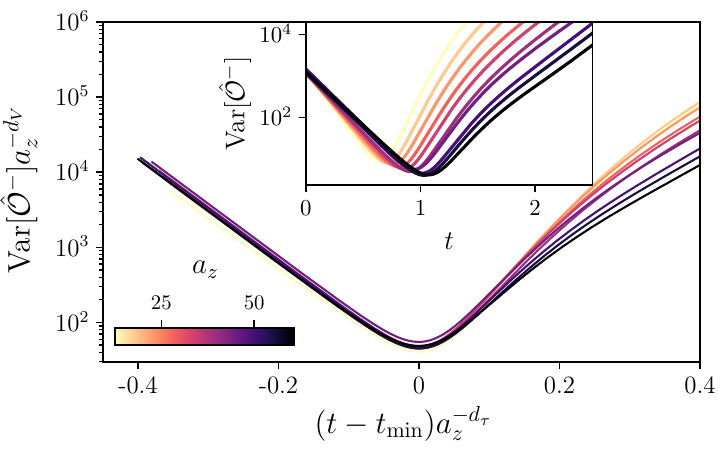} }
\subfloat[$\lambda = 0.8$]{
    \includegraphics[width = 7cm]{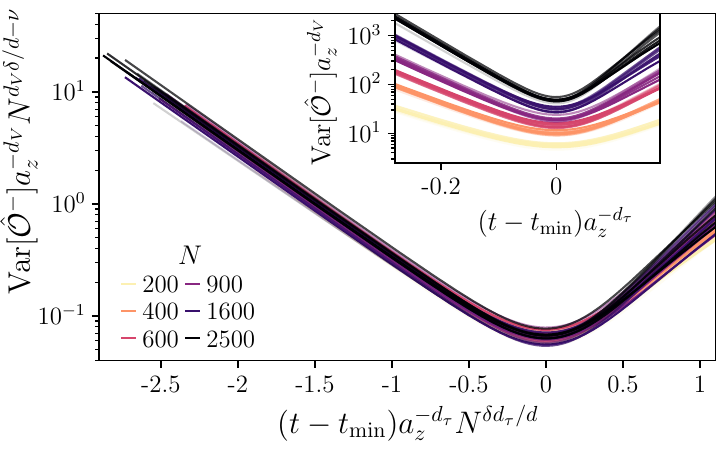} }
    
\subfloat[$\lambda = 1.0$, $N = 2500$]{
    \includegraphics[width = 7cm]{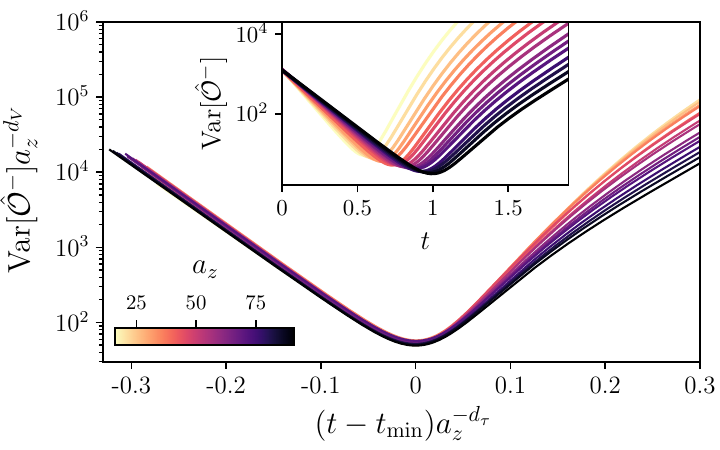} }
\subfloat[$\lambda = 1.0$]{
    \includegraphics[width = 7cm]{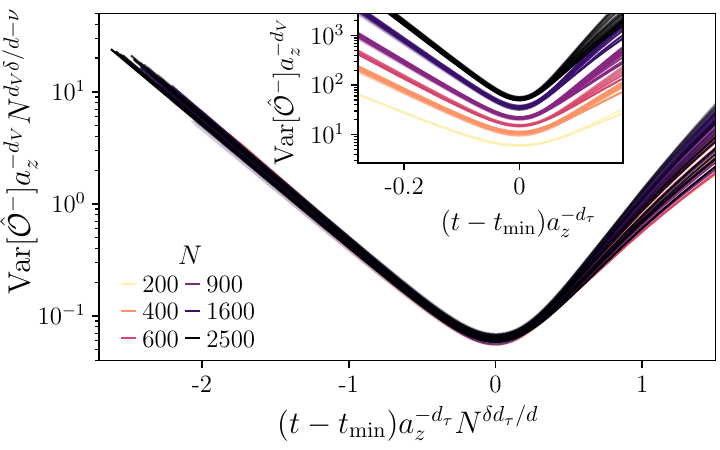} }

\subfloat[$\lambda = 1.4$, $N = 2500$]{
    \includegraphics[width = 7cm]{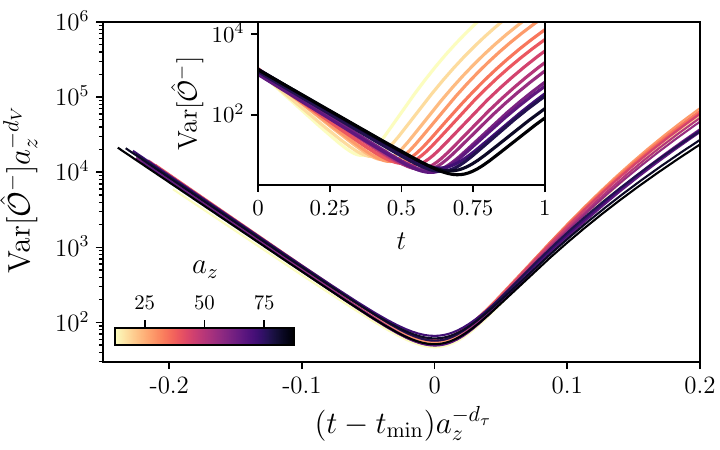} }
\subfloat[$\lambda = 1.4$]{
    \includegraphics[width = 7cm]{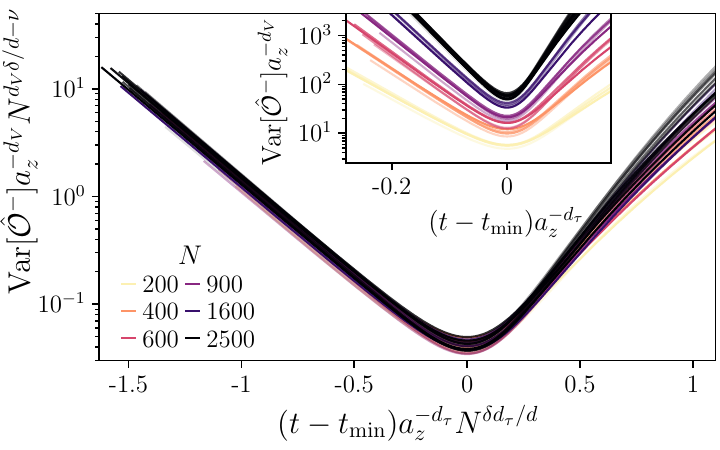} }

\end{center}
    \caption{Raw data and scaling collapses for $\alpha=1$ in $1\mathrm{D}$, for $\lambda = 0.8$  [(a) and (b)], $\lambda = 1.0$  [(c) and (d)] and $\lambda = 1.4$  [(e) and (f)]. Lines for smaller $a_Z$ values are more faded in (b), (d) and (f). The exponents used for the collapse are given in Table~\ref{tab:exponents_lambda}.
    \label{fig:collapse_lambda}}
\end{figure*}

In this appendix we describe the procedure used to extract the critical exponents $(d_V, d_\tau, \nu, \delta)$ entering the scaling ansatz
Eq.~\eqref{eq:scaling_ansatz} of the main text, and present the raw data and scaling collapses underlying Figs.~\ref{fig:Universality_lattice} and \ref{fig:Universality_lambdas}. We follow the optimization procedure introduced in Ref.~\cite{Dresselhaus2022} and used in Ref.~\cite{7g55-lpff}, which defines a cost function to quantify the quality of the scaling collapse.
 
\subsection{Optimization procedure}
 
Given raw data $\{x_{ij}, y_{ij}\}$ for the $j$-th point in the $i$-th data set, ordered by system size $N_i$, we consider the rescaled data
$\{\tilde{x}_{ij}, \tilde{y}_{ij}\} = \{x_{ij} N_i^{d_x},\, y_{ij} N_i^{d_y}\}$ for trial exponents $d_x, d_y$, and rescaled uncertainties
$\tilde{\delta}_{ij} = \delta_{ij} N_i^{d_y}$. A successful scaling collapse requires the rescaled data sets to lie on a common curve. We use
linear interpolation to construct curves $f_i(x)$ and $\delta_i(x)$ from the rescaled data and uncertainties of data set $i$, and define the cost
function~\cite{Dresselhaus2022}
\begin{equation}
  \mathcal{S}(d_x, d_y) \;=\;
  \frac{1}{\mathcal{N}} \sum_{i<k} \sum_{x}
  \frac{[f_i(x) - f_k(x)]^2}{\delta_i(x)^2 + \delta_k(x)^2},
  \label{eq:cost_function}
\end{equation}
where the inner sum runs over the equally spaced grid of $x$ values for which both data sets are defined, and $\mathcal{N}$ is the total number of terms in the double sum. The optimal exponents are those that minimize $\mathcal{S}$. A good collapse is signaled by $\mathcal{S}_{\min} \leq 2$~\cite{Houdayer2004}. Uncertainties on the exponents are estimated as the range over which $\mathcal{S} \leq \mathcal{S}_{\min} + 1$.
 
The numerical uncertainty on individual variance values is estimated from the standard error of the dTWA trajectories. As in Ref.~\cite{7g55-lpff} we use a uniform estimate $\delta \approx 0.04\, \mathrm{Var}[\hat{\mathcal{O}}^-]_{\min}$ for each system size, which we have verified to bound the trajectory-level uncertainty in the cases considered here.
 
The extraction of the four exponents proceeds as follows. The exponent $p$ governing the scaling of the minimal variance with system size is obtained first, from the data underlying Fig.~\ref{fig:p_vs_aZL} of Appendix~\ref{app:identification}, restricted to aspect ratios well inside the partially collective phase. The exponents $d_V$ and $d_\tau$ are then extracted simultaneously by applying the optimization procedure to the largest accessible system size, for which the ansatz reduces to $\mathrm{Var}[\hat{\mathcal{O}}^-]\, a_Z^{-d_V} = f_N[(t-t_{\min})\,
a_Z^{-d_\tau}]$. Given $d_V$ and $d_\tau$, the exponent $\delta$ is obtained from a one-dimensional optimization of the full system-size
collapse, with $\nu$ then determined via the constraint $\nu = p - d_V(1-\delta)/d$ obtained by setting $t=t_{\min}$ in Eq.~\eqref{eq:scaling_ansatz}.
 
\subsection{Raw data and scaling collapses}
 
Figure~\ref{fig:collapse_geom} shows the raw data and scaling collapses underlying Fig.~\ref{fig:Universality_lattice} of the main text, for $\alpha=3$ in $2\mathrm{D}$ on the square, triangular, and honeycomb lattices. The left column shows $\mathrm{Var}[\hat{\mathcal{O}}^-]\, a_Z^{-d_V}$ vs $(t-t_{\min})\, a_Z^{-d_\tau}$ at fixed $N = 10000$ for a range of aspect ratios $a_Z$ in the partially collective phase, demonstrating the collapse achieved by the $a_Z$-rescaling alone. The right column shows the full collapse $\mathrm{Var}[\hat{\mathcal{O}}^-]\, a_Z^{-d_V} N^{d_V\delta/d - \nu}$ vs $(t-t_{\min})\, a_Z^{-d_\tau} N^{\delta d_\tau/d}$ across the available system sizes $N = 2500, 4900, 6400, 8100, 10000$. The exponents extracted for each lattice geometry, together with their uncertainties and the corresponding $\mathcal{S}_{\min}$ values, are reported in Table~\ref{tab:exponents_geom}.
 
Figure~\ref{fig:collapse_lambda} shows the analogous raw data and collapses underlying Fig.~\ref{fig:Universality_lambdas}, for $\alpha=1$ in $1\mathrm{D}$ at the three values $\lambda = 0.8, 1.0, 1.4$ of the interlayer coupling ratio. As for the geometry comparison, the left column shows the partial
collapse at fixed $N=2500$ and the right column the full collapse across system sizes $N = 200, 400, 600, 900, 1600, 2500$. The exponents
extracted at each value of $\lambda$ are reported in Table~\ref{tab:exponents_lambda}.
 
Across all cases the extracted exponents are consistent within their estimated uncertainties (see tables), supporting the universality claim
of Sec.~\ref{sec:universality}. The cost functions $\mathcal{S}_{\min}$ typically lie in the range $\sim 1$--$3$. As noted in Ref.~\cite{7g55-lpff}, these values can be reduced by restricting the optimization to aspect ratios further from the phase boundary, at the cost of reducing the data range available for the extraction.

\end{document}